\documentclass[aps,showpacs,prl,twocolumn,superscriptaddress,apsrev,english]{revtex4-1}
\usepackage{multirow}
\usepackage{color}
\usepackage{amsfonts}
\usepackage{amsmath}
\usepackage{mathrsfs}
\usepackage{amssymb}
\usepackage{wasysym}
\usepackage{graphicx}
\usepackage{babel}
\usepackage{paralist}
\usepackage{indentfirst}
\usepackage{subfigure}
\usepackage{hyperref}
\usepackage{ulem}
\usepackage{soul}
\usepackage{cancel}

\begin{document}
\title{Floquet engineering Hz-Level Rabi Spectra in Shallow Optical Lattice Clock}
\author{Mo-Juan Yin}
\thanks{These authors contributed equally to this work.}
\affiliation{Key Laboratory of Time and Frequency Primary Standards, National Time Service Center, Chinese Academy of Sciences, Xi'an 710600, China}
\affiliation{School of Astronomy and Space Science, University of Chinese Academy of Sciences, Beijing 100049, China}
\author{Xiao-Tong Lu}
\thanks{These authors contributed equally to this work.}
\affiliation{Key Laboratory of Time and Frequency Primary Standards, National Time Service Center, Chinese Academy of Sciences, Xi'an 710600, China}
\author{Ting Li}
\affiliation{Key Laboratory of Time and Frequency Primary Standards, National Time Service Center, Chinese Academy of Sciences, Xi'an 710600, China}
\affiliation{School of Astronomy and Space Science, University of Chinese Academy of Sciences, Beijing 100049, China}
\author{Jing-Jing Xia}
\affiliation{Key Laboratory of Time and Frequency Primary Standards, National Time Service Center, Chinese Academy of Sciences, Xi'an 710600, China}
\author{Tao Wang }
\email{tauwaang@cqu.edu.cn}
\affiliation{Department of Physics, and Center of Quantum Materials and Devices, Chongqing University, Chongqing, 401331, China}
\affiliation{Chongqing Key Laboratory for Strongly Coupled Physics, Chongqing University, Chongqing, 401331, China}
\author{Xue-Feng Zhang }
\email{zhangxf@cqu.edu.cn}
\affiliation{Department of Physics, and Center of Quantum Materials and Devices, Chongqing University, Chongqing, 401331, China}
\affiliation{Chongqing Key Laboratory for Strongly Coupled Physics, Chongqing University, Chongqing, 401331, China}
\author{Hong Chang }
\email{changhong@ntsc.ac.cn}
\affiliation{Key Laboratory of Time and Frequency Primary Standards, National Time Service Center, Chinese Academy of Sciences, Xi'an 710600, China}
\affiliation{School of Astronomy and Space Science, University of Chinese Academy of Sciences, Beijing 100049, China}

\begin{abstract}
Quantum metrology with ultra-high precision usually requires atoms prepared in an ultra-stable environment with well-defined quantum states. Thus, in optical lattice clock systems deep lattice potentials are used to trap ultra-cold atoms. However, decoherence, induced by Raman scattering and higher order light shifts, can significantly be reduced if atomic clocks are realized in shallow optical lattices. On the other hand, in such lattices, tunneling among different sites can cause additional dephasing and strongly broadening of the Rabi spectrum. Here, in our experiment, we periodically drive a shallow $^{87}$Sr optical lattice clock. Counter intuitively, shaking the system can deform the wide broad spectral line into a sharp peak with 5.4Hz line-width. With careful comparison between the theory and experiment, we demonstrate that the Rabi frequency and the Bloch bands can be tuned, simultaneously and independently. Our work not only provides a different idea for quantum metrology, such as building shallow optical lattice clock in outer space, but also paves the way for quantum simulation of new phases of matter by engineering exotic spin orbit couplings.
\end{abstract}
\maketitle
\textit{Introduction.}- Spectroscopy is a fundamental exploratory tool in various fields, including physics, chemistry, astronomy and so on. The line-width of the spectral line directly determines the frequency sensitivity to systematic noise. Optical Lattice Clocks (OLCs) are the most accurate time-frequency measurement devices and even considered as a candidate for defining the unit of time \cite{PrlHK2003,YeScience2008,LudRmp2015,YeNature2021}. An OLC system utilizes the ultra-narrow spectral line provided by the interaction between the ultra-narrow line-width laser and the two-level ultra-cold $^{87}$Sr atoms. In order to weaken the broadening caused by inter-site tunneling process, the atoms are trapped in a deep optical lattice potential \cite{YePRA2009,YeNature2014,YeScience2017}.   
However, the strong trapping potential can introduce higher order light shifts \cite{Wolf2005} and enhance the decoherence caused by Raman scattering \cite{YePRA2011}. Thus, decreasing the inter-site tunneling becomes one of the central tasks for the shallow OLC. The straightforward strategies are: (i) using gravitational potential difference to break translational symmetry \cite{Wolf2005,SL01,SL02}; or (ii) increasing the lattice constant \cite{YePrl2019,Tweezer}. However, the dynamic approach is never considered.

As the conventional approach of the Floquet Engineering (FE), periodically shaking lattice can modify the Bloch-band structure and even suppress the tunneling between nearest-neighbor sites completely \cite{Eckardt2005,Pisa2007,EckardtRMP2017}. Therefore, the FE turns out to be alternative solution for realization of ultra-narrow spectral line in shallow OLC system \cite{ChangCPL2021,ChangPrl2021}. Two major advantages of this approach are: (i) it is naturally suitable for spaceships or satellites, where the gravity is weak or absent \cite{PTPPra2018}; and (ii) it could be easily extended to different geometries, such as three dimensional OLC systems \cite{YeScience2017}. 
\begin{figure}[t]
		\includegraphics[width=0.45\textwidth]{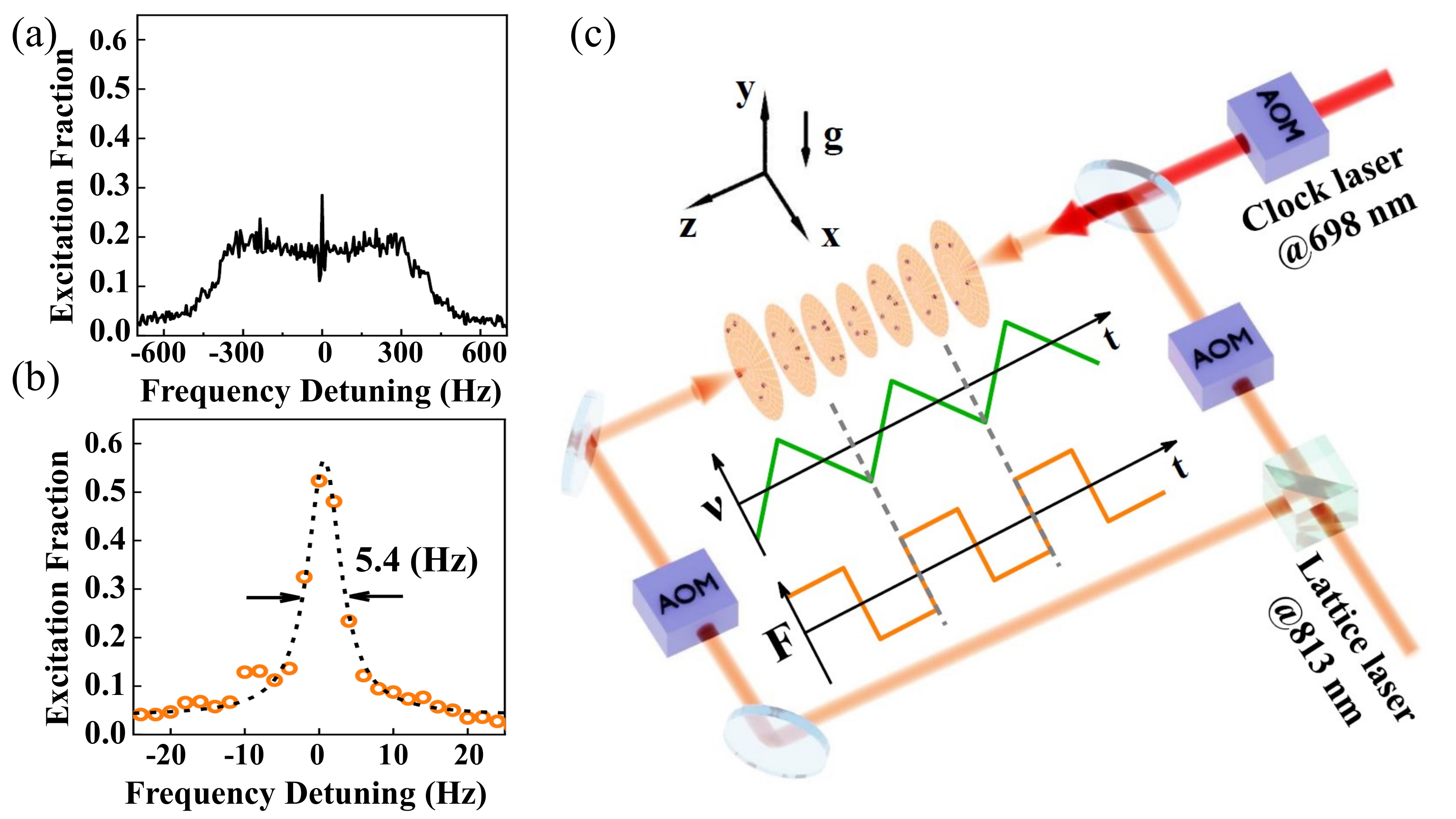}
		\caption{The Rabi spectra of the excited state population at $U_z=9.1$Er, $g_0=21$Hz with probing time $t=150$ms is measured. (a) In the undriven case, the spectrum is weak and broad ($\sim$kHz).(b) In the driven case with frequency excursion of the modulation $\nu_a=13.7644$kHz and driving frequency $\nu_s=600$Hz, a strong and narrow spectral line emerges with 5.4Hz line-width estimated by Lorentz fitting function (dash line). (c) Experimental setup: The $^{87}$Sr atoms are trapped in one dimensional optical lattice formed with two counter-propagating lattice laser beams which are tuned via  an AOM. While the frequency of the stronger lattice laser is locked to the magic wavelength,  the FE is implemented on the weaker one. The AOM aligned on the clock laser is used for band preparation and measurement. The inset presents the effective triangular velocity $v(t)$ of atoms and the effective force $F(t)$ that they feel.}
		\label{fig1}
\end{figure}

In this manuscript, we experimentally implement a Hz-level spectrum in a shallow OLC system. In order to suppress the inter-site tunneling of ultra-cold atoms, we design a FE method implemented on the lattice laser with triangle periodical frequency modulation. Based on the mechanism of dynamical localization \cite{EckardtRMP2017}, the periodic driving can make the atoms pick up an integrated dynamical phase, so that the energy dispersion becomes time-dependent $\epsilon(q(t))$. If the drive is fine-tuned to certain values, the time average dispersion in one period will be $q$ independent (flat-band), which means the hopping process is totally suppressed. As shown in Fig.\ref{fig1} (a-b), a wide spectral line with low excitation fraction is observed for undriven case. In comparison, a sharp spectral line with 5.4Hz line-width can be observed for driven case. The effects of periodic modulation are analyzed with Floquet theory and  verified in the experiment. At last, we test the limitation of our approach.

\textit{Experiment setup}.- We trap an ensemble of nuclear spin-polarized ($m_F=+9/2$) two level $^{87}$Sr atoms in a quasi-one dimensional optical lattice formed by two counter-propagating laser beams originated from unequally splitting one lattice laser at a `magic' wavelength of $\lambda_L= 813.4$nm. As shown in Fig.\ref{fig1} (c), both laser beams are tunable with the help of an acousto-optic modulator (AOM). The stronger one provides a flat trapping potential for $\sim 10^4$ atoms at the temperature $\sim3\mu$K through a global energy shift. This flat potential is not necessary for a three dimensional OLC system\cite{YeScience2017}. The amplitude and frequency of the weak lattice laser can be changed by a voltage variable attenuator (VVA) and an AOM \cite{ChangPrl2021}. After interacting with the clock laser at a wavelength of $\lambda_p=698$nm, the atoms in the ground state $(5s^{2})^{1}$S$_{0} (|g\rangle)$ are excited to the higher energy level $(5s5p)^{3}$P$_{0}(|e\rangle)$ with a long life time ($\sim160$s). When tuning the frequency of the clock laser, we can obtain the Rabi spectra by measuring the excited state population $P_e$ (see supplementary material (SM) \cite{sup} for details).

To induce FE, the frequency of weak lattice laser is periodically driven with a triangle signal function:
\begin{equation}
\Delta \nu (t)=\left\{
\begin{aligned}
&  \nu_a\left(\frac{4t}{T_s}-1\right) \qquad 0\le t< T_s/2  \\
&\nu_a\left(3-\frac{4t}{T_s}\right) \qquad T_s/2\le t< T_s 
\end{aligned}
\right. \label{driven}
\end{equation}
in which $\nu_a$ is the frequency excursion of the modulation, $T_s=1/\nu_s$ is the period with the driving frequency $\nu_s$. In the experiments, $\nu_s$ varies from several hundreds to several kilo-Hertz, so that it can be smaller than the band gap in the axial direction and larger than the hopping amplitudes to avoid parametric heating problem \cite{ht1,ht2,ht3}. Meanwhile, we set $\nu_a$ less than $20$kHz to make the influence of deviation from `magic' wavelength negligible \cite{ChangCPL2021}. As shown in the inset of Fig.\ref{fig1} (c), the periodic driving makes the lattice trap shake with a triangular velocity $v(t)=\lambda_L\Delta \nu(t)/2$. In the co-moving frame of reference, the additional effective force is $F(t)=\frac{M\lambda_L}2\frac{d\Delta \nu(t)}{dt}$ with square form in which $M$ is the mass of $^{87}$Sr atoms. Then based on the Floquet theory \cite{sup}, it is found that the Rabi frequency and Bloch bands can be engineered independently and simultaneously. 
\begin{figure}[t]
	\includegraphics[width=.45\textwidth]{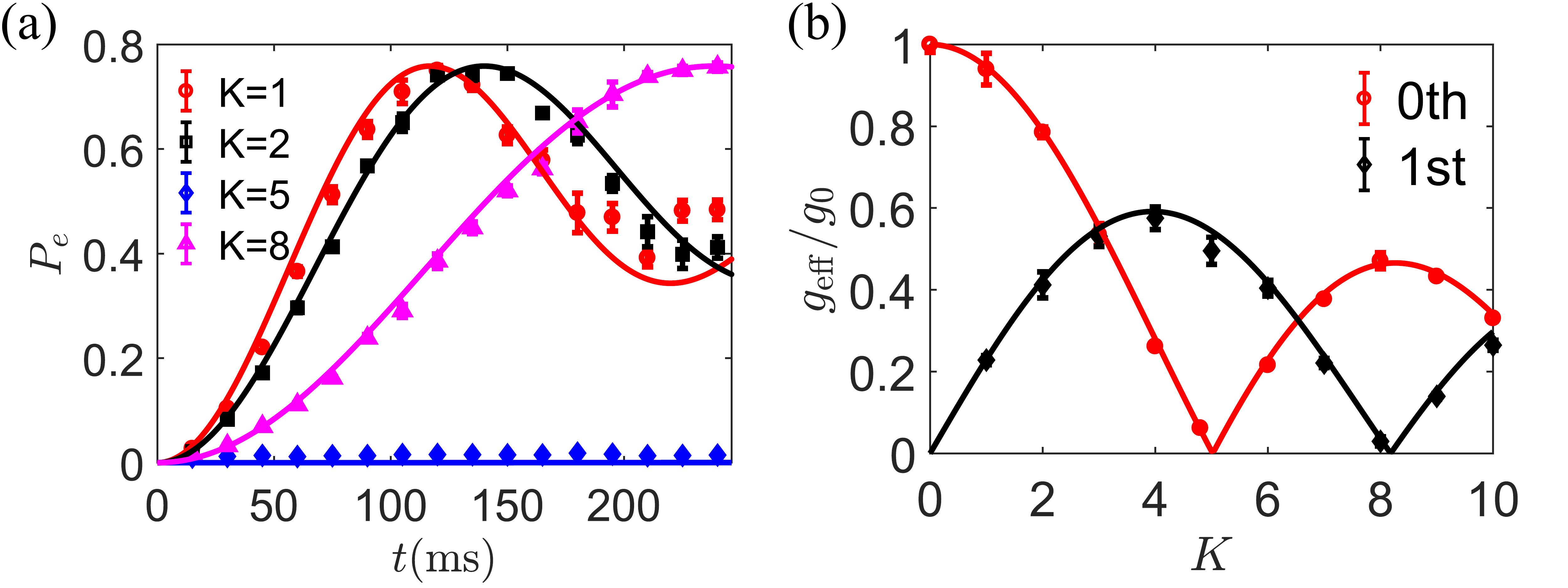}
	\caption{(a) The Rabi oscillation of the zeroth Floquet sideband at different $K$'s. The markers are experimental data, and the solid lines are theoretical results (see Eq.(15) in the SM) with some fitting parameters \cite{ChangCPL2021,sup}. (b) The ratio between the effective Rabi frequency of the Floquet sidebands and its bare value, namely  $g_\mathrm{eff}/g_0$, for both the zeroth and the first order, versus the modulation index $K$ (dots line). The effective Rabi frequency is extracted from the corresponding Rabi oscillations. The markers are the experimental data and the solid lines are the modulation function $\mathcal{R}^{m}(K)$.}
	\label{fig2}
\end{figure}

\textit{Engineering of Rabi frequency.}- In the deep lattice with depth $U_z=90E_r$ (with $E_r=\frac{\hbar^2k_L^2}{2M}\approx3.44$ kHz $\times h$ being the recoil energy), the system can approximately be modeled as two level atoms in identical harmonic traps. In this situation, the Bloch bands barely contribute to the Rabi spectrum and thus the influence of driving on the Rabi frequency can be ignored. 
Since the bare Rabi frequency $g_0$ (definition in SM\cite{sup}) is much smaller than $\nu_s$, we can use the resolved side-band approximation \cite{ChangCPL2021}. The carrier band splits into several Floquet sidebands such that the effective Rabi frequency of the $m$th Floquet sideband is $g_\mathrm{eff}^{m}=g_0\mathcal{R}^{m}(K)$. The modulation function can be explicitly represented as
\begin{eqnarray}
\mathcal{R}^{m}(K)=\int_{-\frac{1}{2}}^{\frac{1}{2}}\cos(m\pi \tilde{t})\cos(\frac{\Phi K}{2}(\tilde{t}^2-\frac{1}{4})+\frac{m\pi}2)d\tilde{t}
\label{RM}
\end{eqnarray}
in which $\tilde{t}=t/T_s$ is the renormalized time, $K=\nu_a/\nu_s$ is the modulation index and the phase $\Phi=\pi\lambda_L/\lambda_p\approx7\pi/6$ is related to the spin orbit coupling (SOC) induced by incommensurate ratio between wavelengths of the lattice and the clock lasers \cite{YeSoc}. Different from conventional Bessel function form \cite{Eckardt2005}, the Eq.(\ref{RM}) contains the Fresnel integrals \cite{sup}.
\begin{figure}[t]
\includegraphics[width=0.45\textwidth]{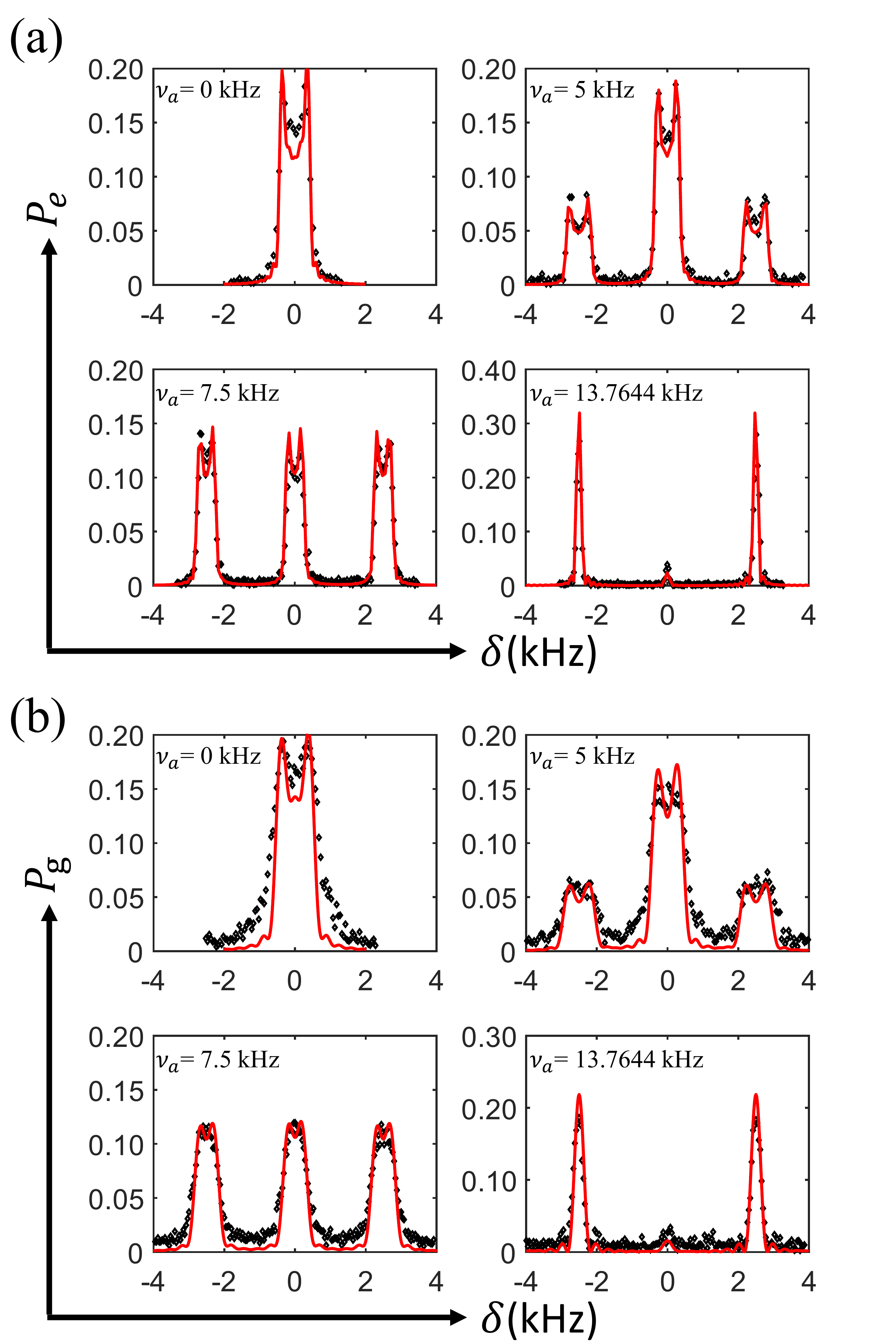}	\caption{(a) The Rabi spectrum of the atoms prepared at the internal state $|g\rangle$ of $n_z=0$ Bloch band. The experimental parameters are $U_z=9E_r$, $g_0=104$Hz, the probe time $t=6$ms, and the driving frequency is fixed at $\nu_s=2.5$kHz. (b) The Rabi spectrum of the atoms prepared at the internal state $|e\rangle$ of $n_z=1$ Bloch band. The experimental parameters are $U_z=27.2E_r$, $g_0=196.7$Hz, the probe time $t=3$ms, and the driving frequency is fixed at $\nu_s=2.5$kHz. The theoretical results (red solid lines), see Eq.(13) in SM, are calculated by taking the same fitting parameters as undriven case except one free fitting parameter radial temperature $T_r$ (see details in \cite{ChangCPL2021,sup}).}
	\label{fig3a}
\end{figure}

In order to extract the effective Rabi frequency, we measure the Rabi oscillations at different values of $K$. As shown in Fig.\ref{fig2} (a), driving the system strongly changes the excitation population. In particular,  at $K=5$ the excitation is even smaller than the  background noise indicating that the effective Rabi frequency is nearly zero. In fact, it is much closer to the zero point of $\mathcal{R}^{0}(K)$, which takes place at $K=5.023$. After extracting $g_\mathrm{eff}^{m}$ and dividing it by $g_0$, the measurement of modulation function can be obtained. Similar to the Rabi oscillation, the experimental results match well with the theoretical predictions, see Fig.\ref{fig2} (b), and the curves for the zeroth and first order Floquet sidebands follow the modulation function $\mathcal{R}^{m}(K)$. The suppressing point $K=8$ of first order Floquet sideband is also consistent with the theoretical prediction $K=8.18$. 

\textit{Engineering the Bloch bands}.- In shallow lattice potentials  the inter-site tunneling of the ultra-cold atoms cannot be ignored. In fact, the hopping amplitude between the sites $l$ and $l'$ can be calculated as $J_{l,l'}^{\vec{n}}$ ($\vec{n}=(n_x,n_y,n_z)$) in harmonic basis $(n_x,n_y)$ in the transverse directions and Wannier basis of band $n_z$ in the longitudinal direction. In this situation, the system can be described by a quantum many body Hamiltonian \cite{sup}. When shaking the lattice, all atoms at different $\vec{n}$ states feel the same effective force, so all the hopping processes are modulated simultaneously. When the driving frequency is much larger than the lattice trap frequencies and tunneling rates, but still much smaller than the band gap, there is no 'Floquet photon assisted hopping' between different bands. Therefore, the hopping amplitudes of all states are dressed by the following normalized sinc function 
\begin{equation}
\mathcal{F}_{l-l'}(\nu_a)=\mathrm{sinc}(\frac{h\nu_a}{4E_r}|l-l'|),
\label{BM}
\end{equation}
where $h$ is the Planck constant. It is quite unusual that the modulation function of bands $\mathcal{F}_{l-l'}(\nu_a)$ only depends on $\nu_a$, while $\mathcal{R}^{m}(K)$ is the function of the modulation index $K$. This indicates that both modulations can be independently tuned by only changing $\nu_a$ and $\nu_s$, which is important for both FE \cite{EckardtRMP2017} and quantum simulation \cite{NRP2020}.

\begin{figure}[t]
	\includegraphics[width=0.45\textwidth]{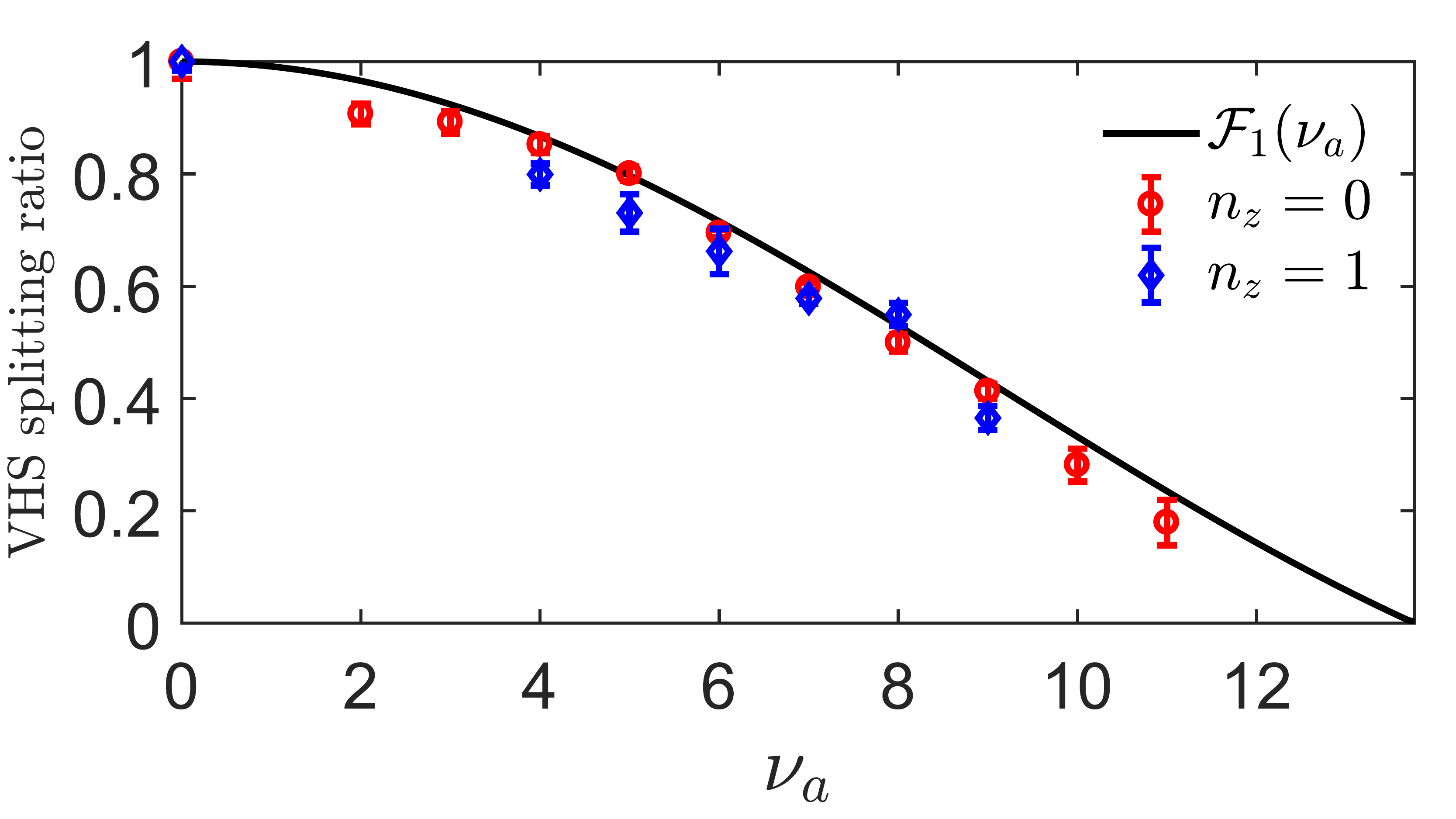}
	\caption{The VHS splitting ratio with $\nu_a$ for both $n_z=0$ and $n_z=1$, which is consistent with modification function $\mathcal{F}_{1}(\nu_a)$.}
	\label{fig3b}
\end{figure}

The atoms in shallow lattice approximately follow the Boltzmann distribution on the Bloch bands labeled with index $(q,\vec{n})$ where $q$ is the momentum. In order to verify the Eq. (\ref{BM}), we prepare all the atoms at $n_z=0$ and $n_z=1$ separately, so the dephasing effect between them can be avoided. In the experiment, the atoms are loaded into a deep shaken lattice potential ($U_z=90E_r$) with target $\nu_a$ and $\nu_s$. Then, after energy filtering \cite{Falke_2014} and adiabatically decreasing the lattice potential to 9$E_r$ in $100ms$, nearly all the atoms are prepared at the internal state $|g\rangle$ of $n_z=0$th Bloch band \cite{sup}. The Fig.\ref{fig3a} (a) shows the Rabi spectrum at a $\pi$ pulse interrogation and driving frequency $\nu_s=$2.5kHz for different $\nu_a$. In the undriven case, the carrier peak is broadened and split to the line shape with double peaks, which is a typical feature of the Van Hove singularity (VHS) resulted from the SOC \cite{YeSoc}. In the presence of driving,  the Floquet sidebands appear with VHS splitting, which becomes smaller by increasing $\nu_a$. At $\nu_a=13.7644$kHz which is close to the zero point solution $\nu_a^0=4E_r/h$ of Eq.\ref{BM}, the carrier peak is almost suppressed accompanying with two high 1st Floquet sidebands with no VHS splitting.
The height variation of the Floquet sidebands can be understood from the modulation of the Rabi frequency. When $\nu_a=13.7644$kHz, the modulation index $K=\nu_a/\nu_s\approxeq5.5$ is quite close to the zero point of $\mathcal{R}^{0}(K)$, so the excitation rate is very weak at zero detuning. On the other hand, considering the nearest neighbor hopping in the tight-banding approximation \cite{YeSoc}, the explicit expression of the splitting between VHS peaks is
\begin{equation}
W_{n_z}(\nu_a)=8\langle J_1^{\vec{n}} \rangle_{T_r}\mathcal{F}_{1}(\nu_a)\sin(\Phi/2),
\end{equation}
where $\langle J_1^{\vec{n}} \rangle_{T_r}$ represents the thermal average of the hopping according to the transverse temperature $T_r$. It is clear that the VHS splitting is irrelevant to the Floquet number and monotonically decreases while increasing $\nu_a$ to $\nu_a^0$.
\begin{figure}[t]
	\centering 
	\includegraphics[width=0.45\textwidth]{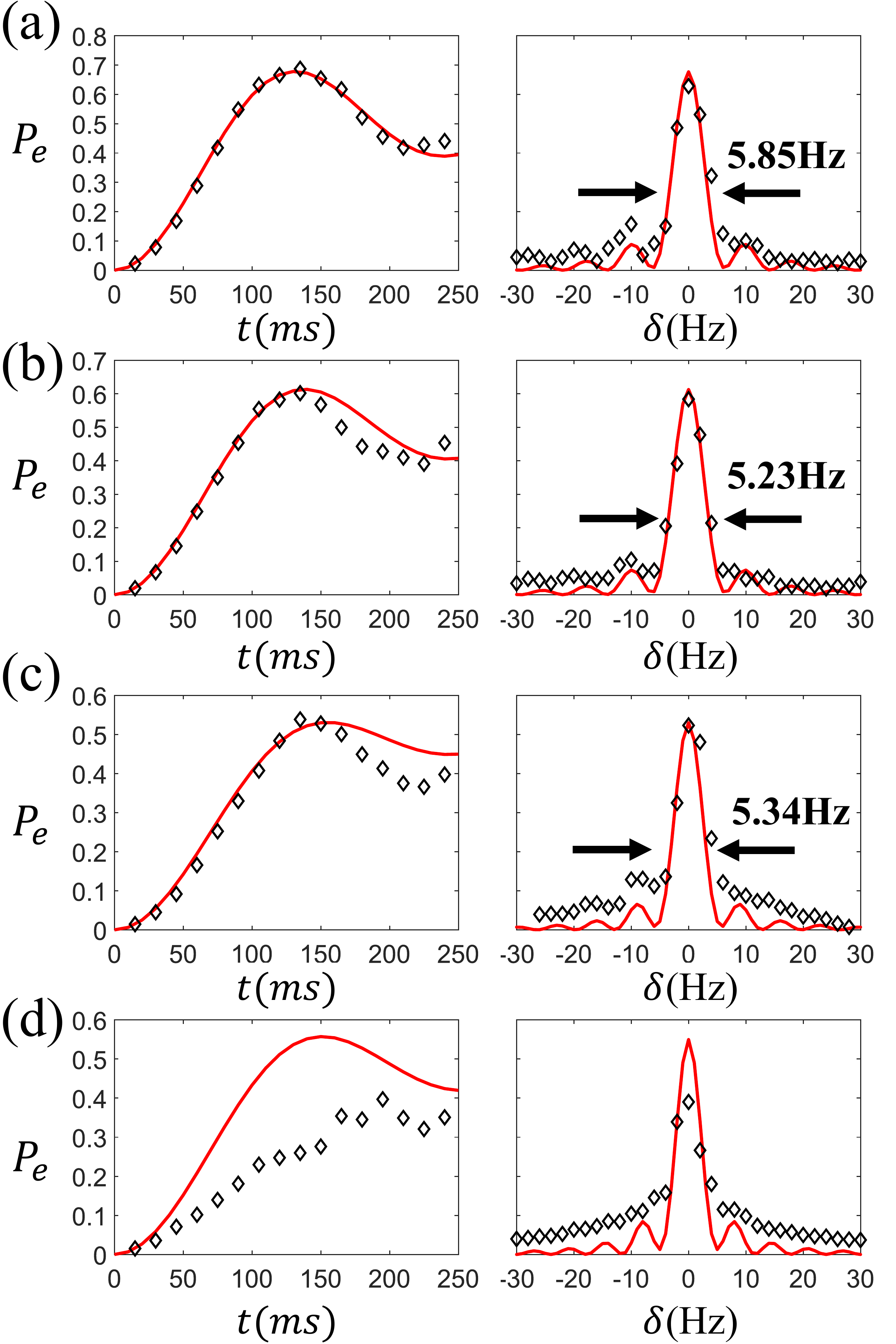}%
	\caption{The experimental data of Rabi oscillations (left) and spectra (right) are measured at $\nu_a=13.7644$kHz, driving frequency $\nu_s=600$Hz and bare Rabi frequency $g_0=21$Hz with different lattice potentials and probe times: (a) $U_z=23E_r$, $t=135$ms; (b) $U_z=15.1E_r$, $t=135$ms; (c) $U_z=9.1E_r$, $t=150$ms; and (d) $U_z=7.2E_r$, $t=150$ms. The theoretical results (see Eq.(13) in SM) of Rabi oscillations (left) and Rabi spectrum (right) are red solid lines. The FWHM of carrier peaks in (a), (b) and (c) are obtained by Lorentz function fitting. The FWHM for $U_z=7.2E_r$ is not given, because its Rabi oscillation is strongly deviated from the theoretical prediction ($\chi^2\approx1.04$ vs $<0.09$ in other cases).} 
	\label{fig4}
\end{figure}

Different from band preparation of $n_z=0$, the target lattice depth for $n_z=1$ Bloch bands is chosen to be deeper $U_z=27.25E_r$, because it can trap more atoms while keeping the hopping amplitude close to $n_z=0$. Besides that, the atoms are prepared in the internal excited state $|e \rangle$, so the Rabi spectrum is obtained by measuring $P_g$ which is the population of the ground state $|g \rangle$. In Fig.\ref{fig3a} (b), we can observe the spectrum of $n_z=1$ presents similar features as $n_z=0$, and it indicates that hopping amplitude of different bands are modulated simultaneously. In order to quantitatively verify the modulation factor of the Bloch bands, we extract the VHS splitting by taking the positions of the VHS peaks. After dividing it by the undriven width, the VHS splitting ratios are plotted in Fig.\ref{fig3b}.
It is clear that both $n_z=1$ and $n_z=0$ data matches well with the theoretical prediction $W_{n_z}(\nu_a)/W_{n_z}(0)=\mathcal{F}_{1}(\nu_a)$. This shows that the Bloch bands are simultaneously modulated with the same function. An immediate implication of this is that one does not need to prepare the system at lower temperatures to suppress the hopping, which will cause severe atom loss.

\textit{Narrow Rabi spectrum}.- In this section, we experimentally demonstrate the Hz-level spectral line. The frequency excursion of the modulation is set to be $\nu_a=13.7644$kHz, so that the hopping amplitude can be suppressed by a factor of $10^{-6}$. Indeed, from the modulation function $\mathcal{F}_{l-l'}(\nu_a)$, one can clearly see that the suppression of the hopping is not limited to the nearest neighbors and all the inter-site hoppings simultaneously go to zero as $\nu_a$ approaching to $\nu_a^0$. That is why we chose the periodic drive to be triangular form instead of the conventional sinusoidal type. Although the next nearest neighbor hopping is much smaller than the nearest neighbor one, the VHS splitting is still tens of Hz for $n_z=1$ when the lattice potential goes down to $15E_r$. Furthermore, due to the long interrogation time, $\nu_s$ should be small to avoid heating from the periodic drive. At the same time, in order to keep the effective Hamiltonian, obtained from high frequency expansion, valid $\nu_s$ should be much larger than the hopping amplitude. To satisfy these conditions, we set the driving frequency to be $\nu_s=600$Hz.

The experimental data of Rabi oscillation and spectrum after a $\pi$ pulse clock laser interrogation for optical lattice potential with various depths are presented in Fig.\ref{fig4}. The Rabi oscillation matches well with the theoretical prediction even at very shallow lattice $U_z=9.1E_r$. Because the $\nu_a$ and $\nu_r$ are not changed, the modulation index $K$ is the same for different lattice potentials. Most importantly, the Hz-level ($\sim$5Hz) Rabi spectra has a high signal-to-noise ratio, and their full width at half maxima (FWHM) is close to the Fourier limit. The theory-experiment consistency of both the Rabi oscillation and the spectral data indicates that the driving under this condition does not bring serious noise or heating problem despite long interrogation time.
 
Certainly, such Floquet engineering still has its limitations. As demonstrated in Fig.\ref{fig4} (d), when the lattice potential goes down to $7.2E_r$, the experimental data starts to deviate from the theoretical prediction, especially the Rabi oscillation. The effective Rabi frequency is apparently much smaller, and the Rabi spectrum have no side-lobes. Such limitation may result from the resonant tunneling effects caused by ``Floquet photons". 

At last, we study the frequency shifts due to the second order Doppler effect, modulation accuracy and deviation of `magic wavelength' \cite{sup}. The preliminary results indicate the periodic driving does not bring serious problems.

\textit{Conclusion}.- We have implemented an ultra-stable periodical modulation method for independent Floquet engineering of the Rabi frequency and the Bloch bands. We have achieved the Hz-level Rabi spectrum in a shallow OLC system which may pave the way for building a Floquet atomic clock.

We thank Abolfazl Bayat a lot for valuable discussions and especially copy-editing. This work is supported by the National Natural Science Foundation of China (Grant No. 61775220), the Key Research Project of Frontier Science of the Chinese Academy of Sciences (Grant No. QYZDB-SSW-JSC004) and the Strategic Priority Research Program of the Chinese Academy of Sciences (Grant No. XDB35010202). X.-F. Z. acknowledges funding from the National Science Foundation of China under Grants  No. 11874094 and No.12047564, Fundamental Research Funds for the Central Universities Grant No. 2021CDJZYJH-003. T. W. is supported by the China Postdoctoral Science Foundation Funded Project (Project No. 2020M673118). 

\bibliographystyle{apsrev4-1}
\bibliography{referen}
\end{document}